\documentclass[a4paper,english]{article}
\usepackage[T1]{fontenc}
\usepackage[latin1]{inputenc}
\usepackage{babel}
\usepackage{graphics}
\usepackage{setspace}
\makeatletter

\providecommand{\LyX}{L\kern-.1667em\lower.25em\hbox{Y}\kern-.125emX\@}
\let\SF@@footnote\footnote
\def\footnote{\ifx\protect\@typeset@protect
    \expandafter\SF@@footnote
  \else
    \expandafter\SF@gobble@opt
  \fi
}
\expandafter\def\csname SF@gobble@opt \endcsname{\@ifnextchar[
  \SF@gobble@twobracket
  \@gobble
}
\edef\SF@gobble@opt{\noexpand\protect
  \expandafter\noexpand\csname SF@gobble@opt \endcsname}
\def\SF@gobble@twobracket[#1]#2{}

 \newcommand{\lyxaddress}[1]{
   \par {\raggedright #1 
   \vspace{1.4em}
   \noindent\par}
 }

\makeatother
\begin{document}
{\raggedleft PF\#2343\par}

{\centering \textbf{\Large Potential model of a 2D Bunsen flame}\Large \par}

{\centering Bruno Denet\par}

\lyxaddress{\centering IRPHE 49 rue Joliot Curie BP 146 Technopole de Chateau
Gombert 13384 Marseille Cedex 13 France}

{\centering submitted to Physics of Fluids\par}

\begin{abstract}
The Michelson Sivashinsky equation, which models the non linear dynamics
of premixed flames, has been recently extended to describe oblique
flames. This approach was extremely successful to describe the behavior
on one side of the flame, but some qualitative effects involving the
interaction of both sides of the front were left unexplained. We use
here a potential flow model, first introduced by Frankel, to study
numerically this configuration. Furthermore, this approach allows
us to provide a physical explanation of the phenomena occuring in
this geometry by means of an electrostatic analogy.



\newpage

\end{abstract}

\section{Introduction \label{sec:introduction}}

The Michelson Sivashinsky equation \cite{siva77} forms the basis
of non linear descriptions of laminar premixed flames. While originally
developed for the case of plane on average flames, variants of this
equation have also been applied to spherical expanding flames \cite{dangelojoulinboury}
\cite{filyandsiva}. However, an extension of this equation valid
for oblique flames has been obtained only recently by Joulin in \cite{searbytruffautjoulin}
by adding a convective term to the original equation , which mimics
the transverse velocity appearing as soon as the flame is maintained
oblique compared to the direction of propagation. The motivation of
this work can be found in the experimental setup of Truffaut and Searby
\cite{truffautsearby}, which the authors have called an inverted
V flame. Actually this configuration is a sort of 2D Bunsen burner
laminar flame, perturbed on one side by an applied electric field.
A comparison of the model equation with the experiment has been remarkably
successful, even from a quantitative point of view, and describes
the development and saturation of wrinkles amplified by the Darrieus-Landau
instability. However, this equation is limited to one side of the
front. Searby and Truffaut have been able to exhibit an effect not
described by the lagrangian Michelson Sivashinsky equation: when the
flame is excited on one side, relatively small cells develop on this
face of the flame but an overall curvature of the flame at large scale
occurs.

This effect can only be explained by a model taking into account both
sides of the flame. It turns out that such a model already exists
in the form of an equation derived by Frankel \cite{frankel}. We
shall describe this model in more detail in section \ref{sec:model},
but for the moment let us just say that the model assumes a potential
flow both ahead and behind the flame, and consists of a boundary integral
equation (the boundary is the flame front, seen as a discontinuity)
involving electrostatic potentials. Some numerical simulations of
this equation can be found in the literature \cite{frankelsiva} \cite{blinnikovsasorov}
\cite{ashurst97} \cite{denetfrankel} in the case of expanding flames,
but it has not been used for the moment in other geometries. We use
here a slightly modified version of this equation to study the 2D
Bunsen flame case. With this method, all the phenomena observed experimentally
are recovered and it is possible to use an electrostatic analogy in
order to get a physical understanding of this problem. 

In section \ref{sec:model} we present the Frankel equation in a form
suitable to the present geometry and we take the opportunity to show
that this equation naturally leads to a qualitative interpretation
of the Darrieus Landau instability. In section \ref{sec:results}
we present the results obtained in the 2D Bunsen flame configuration
for various values of the parameters. Finally section \ref{sec:conclusion}
contains a conclusion.

\section{Model \label{sec:model}}

Let us first introduce some notations. We use two different flame
velocities, the flame velocity relative to premixed gases \( u_{l} \)
and the flame velocity relative to burned gases \( u_{b} \) . Without
gas expansion caused by the exothermic reactions, these two different
velocities would have the same value. However typically the density
in burned gases is five to eight times lower than the density of fresh
gases, which is the main cause of the Darrieus-Landau instability
of premixed flames. If we define \( \rho _{u} \) the density of fresh
gases, \( \rho _{b} \) the density of burnt gases, \( \gamma =\frac{\rho _{u}-\rho _{b}}{\rho _{u}} \)a
parameter measuring gas expansion (\( \gamma =0 \) without exothermic
reactions), then \( u_{b}=\frac{u_{l}}{1-\gamma } \)because of mass
conservation. In many articles, the notion of flame velocity relative
to burned gases is never used, however in the original Frankel paper,
what is called flame velocity is actually \( u_{b} \). We obtain
below a Frankel equation relative to fresh gases, and this form of
the equation will be simulated in the following section. The derivation
closely parallels the original one, except for the flame velocity
used and for the geometry . A sketch of the configuration can be found
in Figure \ref{fig: configuration}, the unburnt gases are injected
at the velocity \( U \). The flame has a shape typical of a Bunsen
burner and propagates normally at a velocity \( u_{l} \) in the direction
of unburnt gases. We also consider that \( U \) is constant in space
and time and that the flame is attached at two constant position points. 

The idea behind the Frankel equation is the following: the Michelson
Sivashinsky equation is obtained as a development with \( \gamma  \)
as an expansion parameter. It has been shown in \cite{sivaclavin}
that at the lowest order in \( \gamma , \) the equation obtained
by neglecting vorticity reduces to the Michelson Sivashinsky equation.
So let us neglect vorticity everywhere, including in the burnt gases,
we can define a velocity potential (\( w_{u} \) and \( w_{b} \)
in the fresh and burnt gases), which is solution of the 2D Laplace
equation: 

\[
w_{xx}+w_{yy}=0\]

On the flame, which is a discontinuity in this formulation, the velocity
potential has to satisfy 

\[
w_{u}=w_{b}\]

\[
\left( -\frac{\partial w_{b}}{\partial n}+V-\overrightarrow{U}.\overrightarrow{n}\right) \rho _{b}=\left( -\frac{\partial w_{u}}{\partial n}+V-\overrightarrow{U}.\overrightarrow{n}\right) \rho _{u}\]

\[
-\frac{\partial w_{u}}{\partial n}+V-\overrightarrow{U}.\overrightarrow{n}=u_{l}+\varepsilon \kappa \]

\( \kappa  \) is the curvature at a given point on the flame , \( \varepsilon  \)
is a contant number proportional to the Markstein length, \( \overrightarrow{n} \)
is the normal vector at the current point on the front, in the direction
of propagation. After some calculations, an evolution equation is
obtained, valid for an arbitrary shape of the front (the reader is
referred to \cite{frankel} for more details on the derivation, please
remember that the flame velocity used in this paper is \( u_{b} \)
).

\begin{equation}
\label{mathed:V}
V(\overrightarrow{r},t)=u_{l}+\varepsilon \kappa +\overrightarrow{U}.\overrightarrow{n}+\frac{1}{2}\frac{\gamma }{1-\gamma }u_{l}-\frac{u_{l}}{2\pi }\frac{\gamma }{1-\gamma }\int _{S}\frac{\left( \overrightarrow{\xi }-\overrightarrow{r}\right) .\overrightarrow{n}}{\left| \overrightarrow{\xi }-\overrightarrow{r}\right| ^{2}}dl_{\xi }+\overrightarrow{V}_{boundary}.\overrightarrow{n}
\end{equation}

This equation gives the value of the normal velocity \( V \) on the
front as a sum of several terms, the laminar flame velocity with curvature
corrections, the velocity of the incoming velocity field and an induced
velocity field (all the terms where \( \gamma  \) appears) which
contains an integral over the whole shape (indicated by the subscript
\( S \) in the integral ). This integral is a sum of electrostatic
potentials. 

Let us recall that, as is well-known, the formula for the induced
velocity field at a position not located on the front is given by
a different formula, which we shall use to reconstruct the velocity
everywhere once the shape is known. 

\begin{equation}
\label{mathed:Vinduced}
\overrightarrow{V}_{induced}(\overrightarrow{r},t)=-\frac{u_{l}}{2\pi }\frac{\gamma }{1-\gamma }\int _{S}\frac{\left( \overrightarrow{\xi }-\overrightarrow{r}\right) }{\left| \overrightarrow{\xi }-\overrightarrow{r}\right| ^{2}}dl_{\xi }
\end{equation}

i.e. compared to equation (\ref{mathed:V}) the induced velocity term
does not contain the constant term \( \frac{1}{2}\frac{\gamma }{1-\gamma }u_{l} \)
. The last term \( V_{boundary} \) is a potential velocity field
(continuous across the flame) added to the equation in order to satisfy
the boundary conditions. Here the condition is simply that

\[
\left( \overrightarrow{V}_{induced}+\overrightarrow{V}_{boundary}\right) .\overrightarrow{n}=0\]

at the injection location, where \( \overrightarrow{n} \) is parallel
to \( \overrightarrow{U} \) , so that \( V_{boundary} \) is given
by the same type of integral as \( V_{induced} \) , but over the
image of the front, drawn as a dashed line in Figure \ref{fig: configuration}. 

Naturally, the shape evolves according to the velocity \( V(\overrightarrow{r},t) \)
:

\[
\frac{d\overrightarrow{r}}{dt}=V\overrightarrow{n}\]

where \( \overrightarrow{r} \) denotes the position of the current
point of the front.

We would like at this point to emphasize the analogy between the flame
propagation problem and electrostatics. Let us consider a plane flame,
infinite in the transverse direction. If we inject fresh gases with
a velocity equal to \( u_{l} \), then the flame does not advance
and the velocity in the burnt gases is \( u_{b} \)(left of Figure
\ref{fig:plan}). As explained before we can add any potential velocity
field which does not generate a jump of velocity across the flame
(which is already described by \( V_{induced} \)) so as to satisfy
boudary conditions. In particular we can add a constant velocity field,
which would show that if the velocity field in fresh gases is zero,
the flame propagates at \( u_{l} \), if the velocity field in the
burnt gases is zero, the apparent flame propagation velocity is \( u_{b} \).
However equation (\ref{mathed:V}) corresponds to the symmetrical
situation depicted in the middle of Figure \ref{fig:plan}: the velocity
field in the burnt and fresh gases has the same value (in the opposite
direction) \( \frac{u_{b}-u_{l}}{2}=\frac{u_{l}}{2}\frac{\gamma }{1-\gamma } \)
(the constant value that appears in equation (\ref{mathed:V})) and
the flame propagates at the apparent velocity \( \frac{u_{b}+u_{l}}{2} \).
There is an analogy of this situation with a uniformly charged infinite
plane in electrostatics (Figure \ref{fig:plan}, right), which generates
on both side an electric field of value \( \frac{\sigma }{2\varepsilon _{0}} \)
in the international system of units. 

One of the purposes of this paper is to show that this analogy enables
us to have a physical understanding of phenomena occurring in unstable
premixed flames. Let us start in this section by showing that we can
explain qualitatively the Darrieus-Landau instability. We consider
once again an infinite plane flame. The induced velocity \( \frac{u_{b}-u_{l}}{2} \)just
above the flame is simply obtained by integration over the whole front
(left of Figure \ref{fig:darrieuslandau}). Just on the front , the
integral term in equation (\ref{mathed:V}) vanishes because of symmetry
reasons so that this term\( \frac{u_{b}-u_{l}}{2} \) has to be added
explicitely in (\ref{mathed:V}). Let us consider now a wrinkled flame
(right of Figure \ref{fig:darrieuslandau}). The induced velocity
very close to the flame is obtained as before by integration over
the whole front. However now we can see in the figure (point A) that
a part of the integration produces a velocity in the direction opposite
to the previous induced velocity field, which tends to amplify the
existing wrinkle. A similar reasoning could be performed at point
B, showing also an amplification. Furthermore, it is easily seen that
smaller wavelengths lead to a higher instability, a known property
of the Darrieus Landau instability when curvature effects are neglected.
In conclusion of this paragraph, we can see that the potential approximation
leads to a physical explanation of the instability, which would be
much more difficult to achieve for the complete problem. 

Before presenting the results, we can note that an approach equivalent
to the Frankel equation was used in \cite{pinderatalbot86} \cite{ashurst87}
\cite{pinderatalbot88} \cite{rheetalbotsethian}; The idea leading
to this equation was slightly different. In the Frankel case the idea
was to generalize the Michelson Sivashinsky equation which had a lot
of success for plane on average flames. Pindera and Talbot wanted
to provide a complete numerical solution of the flame problem when
the flame is seen as a discontinuity. When baroclinity is neglected
we obtain exactly the Frankel equation problem, and it could be supplemented
by resolutions with vortex methods in order to have a rigorous description
of the velocity field (with creation of vorticity in the burnt gases).
But this resolution is not very easy (i.e. not easier than a direct
numerical resolution of the problem). In this article, in the spirit
of the Michelson-Sivashinsky equation, and as in Ashurst's work (see
for instance \cite{ashurst97}), we keep the potential approximation
for the 2D Bunsen flame case described in the introduction, and show
that this description is qualitatively correct.

\section{Results\label{sec:results}}

In the Truffaut-Searby configuration that we try to model here, the
2D flame is perturbed on one side, close to the base of the flame,
by an electrostatic apparatus. As a result, 2D wrinkles are created,
with a wavelength depending on the frequency of the applied electric
field, and propagate along the flame because of the tangential velocity.
During their trip from the base to the tip of the flame, the perturbations
are amplified by the Darrieus-Landau instability. A photograph of
an experiment is given in Figure \ref{fig:experiment}. The displacement
of the cells during the exposure time can be seen, and gives an idea
of the dynamics of the flame. A successful description of the phenomena
described above has been given by a Lagrangian Michelson-Sivashinsky
equation in \cite{searbytruffautjoulin}. However this approach is
inherently limited to one side of the flame, and cannot describe phenomena
involving interaction of both sides, such as the large scale curvature
of the flame that can be observed in Figure \ref{fig:experiment}.

We use here the Frankel equation {[}\ref{mathed:V}{]} to study this
problem. Let us first verify that we are able to recover the experimental
results with this model. We excite the flame by applying a velocity
\( v_{x}=a\cos (\omega t) \) at the fourth point starting from the
bottom of the flame, \( x \) is the direction perpendicular to the
injection velocity. As the flame front is described by a chain of
markers (technical details on the numerical method can be found in
\cite{denetfrankel}) the position of this point is not strictly constant,
but the perturbations being small at this location, we have found
that this type of forcing is satisfactory i.e. it generates a well
defined wavelength related to the frequency and the tangential velocity.
Figure \ref{fig:fleches} is obtained by a numerical simulation for
parameters \( U=10 \) \( L=4 \) \( \varepsilon =0.2 \) \( a=1 \)
\( \gamma =0.8 \) \( \omega =50 \) (\( L \) is the distance at
the base of the flame, in all the calculations \( u_{l}=1 \)) . The
number of markers used in the simulation is not constant in time but
is typically around 900. On this figure are plotted both the shape
of the front and the induced velocity field (actually \( V_{induced}+V_{boundary} \)
, see equation (\ref{mathed:Vinduced})). Arrows very close to the
front are not drawn in this figure, as the formula for the induced
velocity field cannot be applied when the distance from the front
is of the order of the distance between successive markers. Naturally
the analogy with Figure \ref{fig:experiment} is striking: both the
wrinkle amplification and subsequent saturation by non linear effects
are observed, but also the large scale curvature (i.e. as we get closer
to the tip, the side opposite to the forcing gets more and more deflected).
This progressive deviation of the right side of the flame is a consequence
of the induced velocity field, which has a component towards the right
when one approaches the tip, as can be seen in Figure \ref{fig:fleches}.

This large scale deviation was observed by Truffaut and Searby, and
for the moment the Frankel equation succeeds in producing this effect.
But a qualitative interpretation can also be obtained. We have seen
that it can be considered that the induced velocity is caused by a
uniformly positively charged front (the electrostatic analogy). As
the wrinkle develops when we get closer to the tip, the velocity has
a sinusoidal component with the wrinkle wavelength in the direction
parallel to the front, and, as a solution of a Laplace equation, this
component decays exponentially in the perpendicular direction. So,
the perturbation with the wrinkle wavelength will be small on the
side opposite to the forcing. Sufficiently far, the perturbed side
can just be considered as a straight unperturbed line, but with a
charge higher than before (a consequence of the Gauss theorem: the
charge inside a small rectangle is higher because of the wrinkle).
So the situation is close to the sketch of Figure \ref{fig:ondule}.
At the base of the flame, the charge is the same for both sides of
the flame. As a result the velocities induced by both sides of the
flame have the same absolute value, but are in the opposite direction
if we consider for simplicity both sides parallel. The total induced
velocity field nearly cancels in the fresh gases, and is high in the
burnt gases. On the contrary, in a zone with well-developed wrinkles,
the charge is higher on the wrinkled side because of the previous
argument, and generates a velocity with a higher absolute value. The
total induced velocity field is thus directed towards the right, and
tends to cause a deviation of both sides of the flame (see Figure
\ref{fig:fleches}). There is also the fact that the front cannot
be considered infinite close to the tip, which generates a velocity
upward because there is no compensation of the upward velocity field
created by the charges below (on the other hand, at the base of the
flame the velocity is very small because the downward \( V_{induced} \)
is compensated by \( V_{boundary} \) which is a velocity field created
by the image of the flame: see Figure \ref{fig: configuration}). 

The potential flow model is thus in very good qualitative agreement
with the experiments of Searby and Truffaut. However, it does not
seem possible to obtain the same quantitative agreement for the development
of a wrinkle on one side, as in the lagrangian Michelson Sivashinsky
case \cite{searbytruffautjoulin}. The reason can be understood in
the following way : actually the results of \cite{searbytruffautjoulin}
are obtained by a Michelson Sivashinsky equation with modified coefficients.
The dispersion relation is fitted in order to be in agreement with
the experimental results, then there is also a modification (compatible
with an expansion in \( \gamma  \)) of the coefficient of the non
linear term. With these modifications the development of perturbations
along the front can be described quantitatively. However in our case
a modification of the coefficients to fit the dispersion relation
would have also an effect on the other side on the front. It seems
unlikely that the same set of coefficients can describe precisely
both the dispersion relation and the effect of one side on the other,
although some kind of compromise can perhaps be found. 

So we will limit ourselves in this paper to qualitative results obtained
by the Frankel equation. A positive point of this model is that we
can vary easily the physical parameters, contrary to experiments.
Of course changing the width at the base of the flame involves a whole
new burner, but it is also difficult experimentally to increase the
injection velocity, because the flame has to be anchored on the rod
where the electric field is applied. 

When perturbations of the 2D Bunsen flame do exist on both sides of
the front, the question arises of knowing the type of modes that will
develop, sinuous or varicous. It is possible to impose one of this
mode by applying an electric field on both sides, with a well-defined
phase relationship, but we prefer here to study what will happen naturally.
We apply now a white noise at the base of the flame, at the same location
as before, but on both sides. A typical front shape, for a small width
\( L=1 \) and a large injection velocity \( U=40 \) is shown in
Figure \ref{fig:uinj40aa1g02abblanc5exp08}. The other parameters
are \( \varepsilon =0.2 \) and \( \gamma =0.8 \), the amplitude
of the white noise being \( a=5 \), \( v_{x}=a(random-0.5) \) where
\( random \) is a random number (uniform distribution between \( 0 \)
and \( 1 \)), always imposed on the fourth point from the base of
the flame .The number of points used in the simulation is typically
7000. The main conclusion of different calculations, which can be
seen on the Figure, is the following: in a first stage, perturbations
develop on both sides in an independent way, neither the sinuous nor
the varicous mode are favored. However, for a sufficient length of
the flame, close to the tip, the sinuous mode is the dominant one.
The sinuous zone corresponds to a distance between both sides of the
order of the wavelength, which is natural for a potential model. Actually,
when this distance is comparable to the wavelength, the perturbations
have the following choice: be damped because they do not have a sufficient
distance to develop, or amplify as before but in a sinuous mode. A
similar sinuous mode has been obtained in \cite{joulinsiva92} for
twin flames in stagnation point flows.

Another interesting problem is the wavelength itself. For a planar
on average flame without gravity, perturbations with the most amplified
wavelength emerge from a flat front, then non linear effects come
into play, cells merge and in the end only one cell remains. This
effect is generally not observed in oblique flames, simply because
the available length is too short, and the wrinkles reach the tip
before merging. In order to observe the merging (in a lagrangian way)
we consider a very large flame: \( U=10 \) \( L=20 \) \( \varepsilon =0.2 \)
\( \gamma =0.8 \). These conditions correspond either to a very large
flame at atmospheric pressure or to a flame at high pressure. We start
the simulation from a flame which has been submitted for some time
to a white noise everywhere, not only at the base, as in Figure \ref{fig:uinj40aa1g02abblanc5exp08}.
This front can be seen in Figure \ref{fig:uinj10aa20g02abblanc5exp08}
which will be used as an initial condition. The small cells of this
flame evolve in the Figures \ref{fig:uinj10aa20g02abblanc0suiteexp08}
to \ref{fig:uinj10aa20g02abblanc0suite5exp08} without any noise.
The merging of the cells, similar to the one observed in planar on
average flame, appears, but occurs here in a lagrangian way, as the
cells are convected towards the tip. The reader can find a similar
behavior obtained recently with the lagrangian Michelson Sivashinsky
equation in \cite{bouryjoulin.ctm2002} (see also the corresponding
animation which can be found on the Combustion Theory and Modelling
web site). The effect seen previously exists also here, sinuous modes
dominate at the tip. However as the perturbations develop, the overall
surface being more or less constant, the height of the flame becomes
smaller, as seen in Figure \ref{fig:uinj10aa20g02abblanc0suite5exp08}.
The same effect is observed in turbulent flows, and as the available
length on each side of the mean flame is smaller, the merging of cells
becomes more difficult. An example of turbulent flames obtained in
this configuration, which are relatively similar to the solutions
obtained here in the presence of noise, can be found in \cite{Kobayashiwilliams}.

After this solution, we have not seen a continuation of the merging
process, on the contrary new cells appear on the front while some
cells merge. Also we have seen no sign that the flame will ultimately
recover its unperturbed shape. The two last observations are related
to the level of numerical noise present in the simulation, for instance
insertion and deletion of markers. Actually we have found in another
oblique configuration (V flame with initial perturbations) and for
similar sizes, that it is possible to recover a more or less stationary
flame by using twice as many markers in the simulation. In \cite{searbytruffautjoulin},
it has been suggested that for a sufficient injection velocity (i.e.
in normal situations), the instability of the flame is convective,
which seems to be the case with the potential model used here. However,
just as in the expanding flame case, it appears that for large sizes,
the flame is extremely sensitive to any external noise. Very small
injection velocities, corresponding to an absolute instability, could
actually lead to flashback. We have seen some indications that this
phenomenon actually occurs, but it was not possible to describe correctly
the evolution with the current algorithm after the flame enters the
tube.

\section{Conclusion \label{sec:conclusion}}

In this article, we have studied the 2D Bunsen flame configuration
proposed by Truffaut and Searby by means of a model equation which
considers the flow as potential both ahead and behind the flame. It
has been shown that this approximation gives a good qualitative description
of the phenomena observed. On the other hand, it is probably difficult
to obtain a complete quantitative agreement with experiments with
this model, contrary to modified versions of the Michelson-Sivashinsky
equation, which do not describe the repulsion effect of the perturbed
side of the front. We expect however that other oblique flame geometries
can be studied with this approach, such as V flames or 3D premixed
Bunsen burner flames. However the 3D case is very difficult technically,
because the treatment of reconnections implemented here in 2D is challenging
in 3D (it is ironic that these reconnections occurring close to the
tip play a minor role in the physics, except for symmetrical forcings,
but are the main numerical difficulty of the problem). Another natural
extension of this work would be to consider a flame in a turbulent
flow, in order to get an estimate of the relative importance of turbulence
and of the Darrieus-Landau instability. The author has already done
some work with turbulence and hydrodynamic instability for expanding
flames, but this could now be extended to oblique flames. We have
also found in this article that the qualitative behavior can be different
for flames in different geometries. This effect is known in laminar
flame configurations, but certainly deserves attention also in the
turbulent case.

\textbf{Acknowledgments}: the author would like to thank J.M. Truffaut,
G. Searby and G. Joulin for helpful discussions, and for the photograph
included in this article.

\bibliographystyle{unsrt}
\bibliography{turb2d.reflib}

\section*{List of Figures}

\begin{figure}[p]

\caption{\label{fig: configuration} Denet, Phys. Fluids\protect \\
}

{\centering \resizebox*{!}{0.75\textheight}{\includegraphics{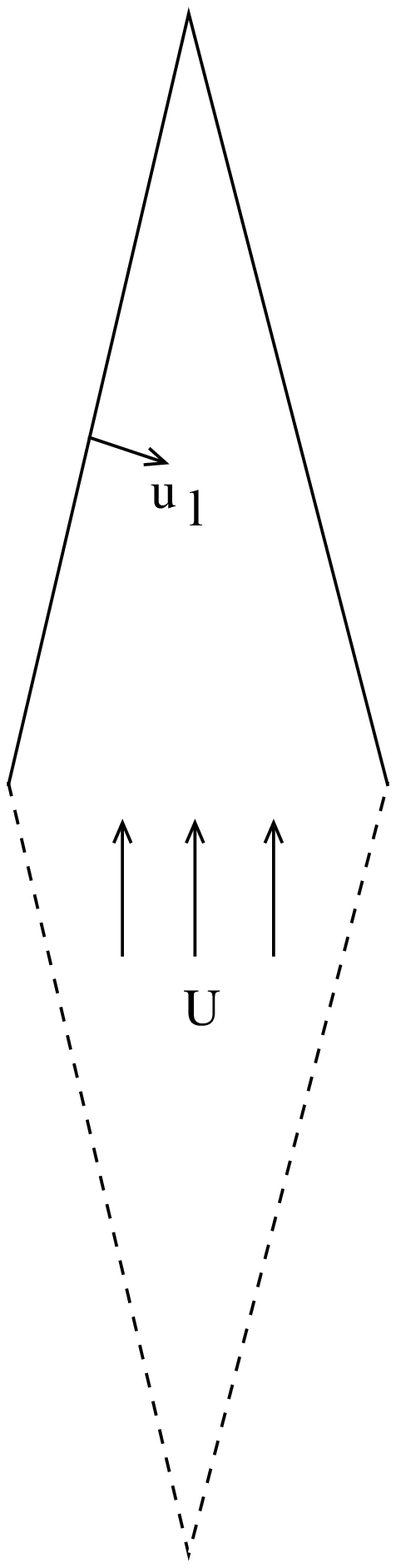}} \par}
\end{figure}

Figure \ref{fig: configuration} : Configuration. Solid line: flame
front, dashed line: electrostatic image of the front 

\begin{figure}[p]

\caption{\label{fig:plan} Denet, Phys. Fluids\protect \\
}

{\centering \resizebox*{0.75\textwidth}{!}{\includegraphics{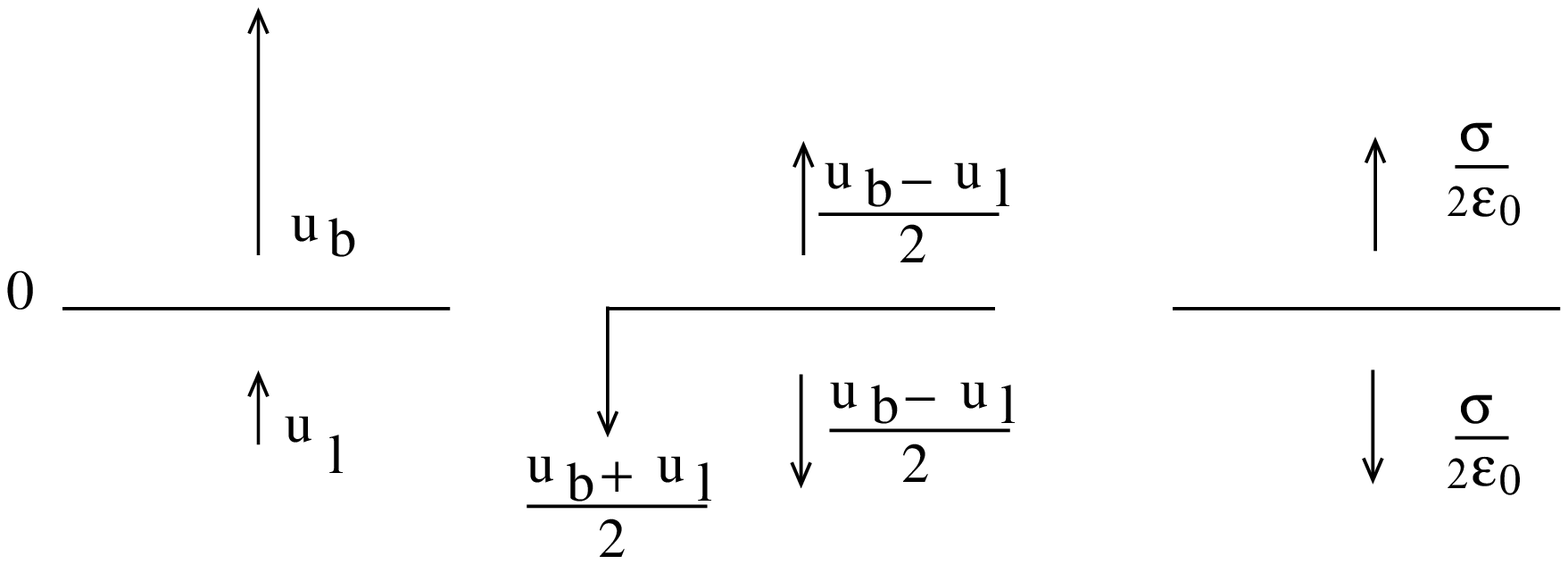}} \par}

\vspace{15cm}
\end{figure}

Figure \ref{fig:plan}: Left: flame stabilized because of the velocity
\( u_{l} \) in the fresh gases. Middle: flame seen in a reference
frame with symmetrical velocities in burnt and fresh gases. Right:
electrostatic analogy with a uniformly charged plane

\begin{figure}[p]

\caption{\label{fig:darrieuslandau} Denet, Phys. Fluids\protect \\
}

{\centering \resizebox*{0.75\textwidth}{!}{\includegraphics{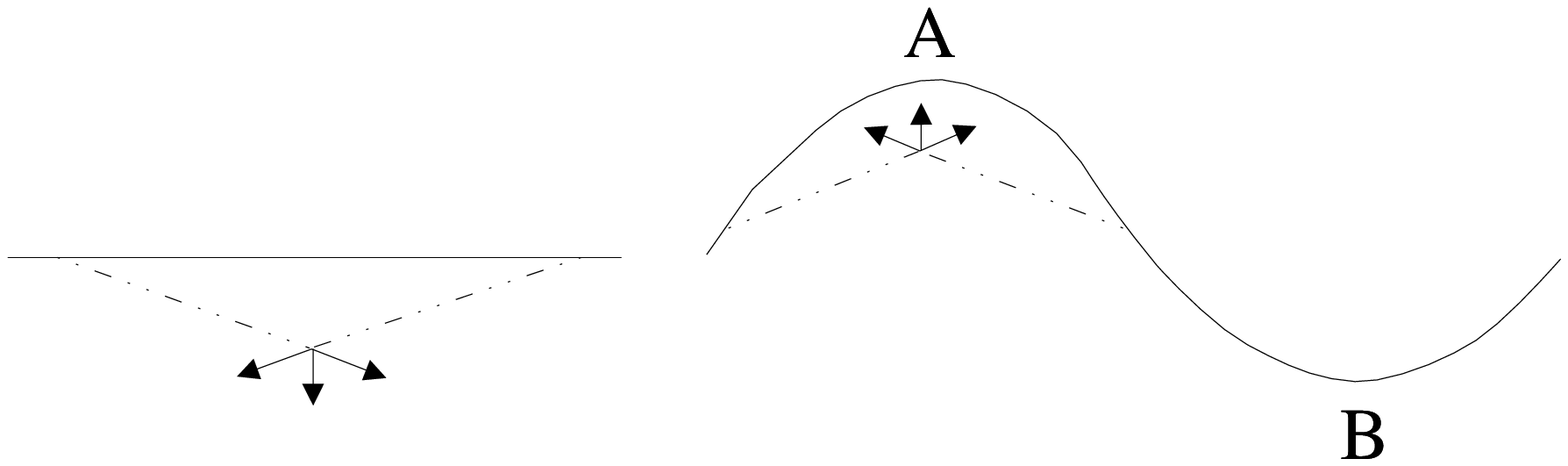}} \par}
\end{figure}

Figure \ref{fig:darrieuslandau}: A qualitative explanation of the
Darrieus-Landau instability : comparison of a plane and a wrinkled
flame. Effect of nearby points on the propagation velocity.

\begin{figure}[p]

\caption{\label{fig:experiment} Denet, Phys. Fluids \protect \\
}

{\centering \resizebox*{!}{0.9\textheight}{\rotatebox{0}{\includegraphics{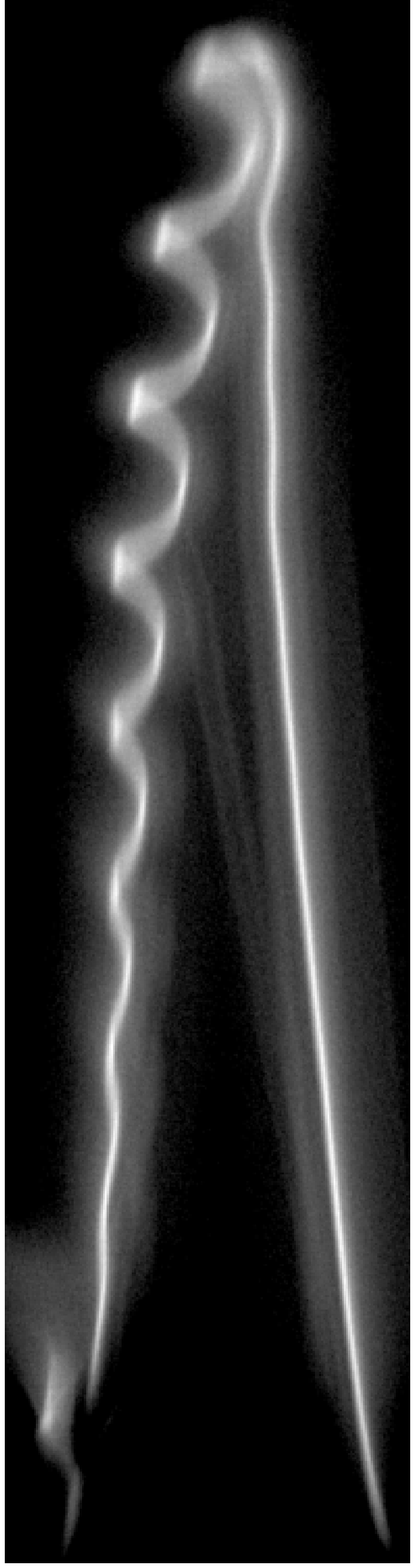}}} \par}
\end{figure}

Figure \ref{fig:experiment}: Photograph of an experimental 2D Bunsen
flame submitted to a sinusoidal forcing on one side (courtesy of J.M
Truffaut and G. Searby)

\begin{figure}[p]

\caption{\label{fig:fleches} Denet, Phys. Fluids\protect \\
}

{\centering \resizebox*{!}{0.9\textheight}{\includegraphics{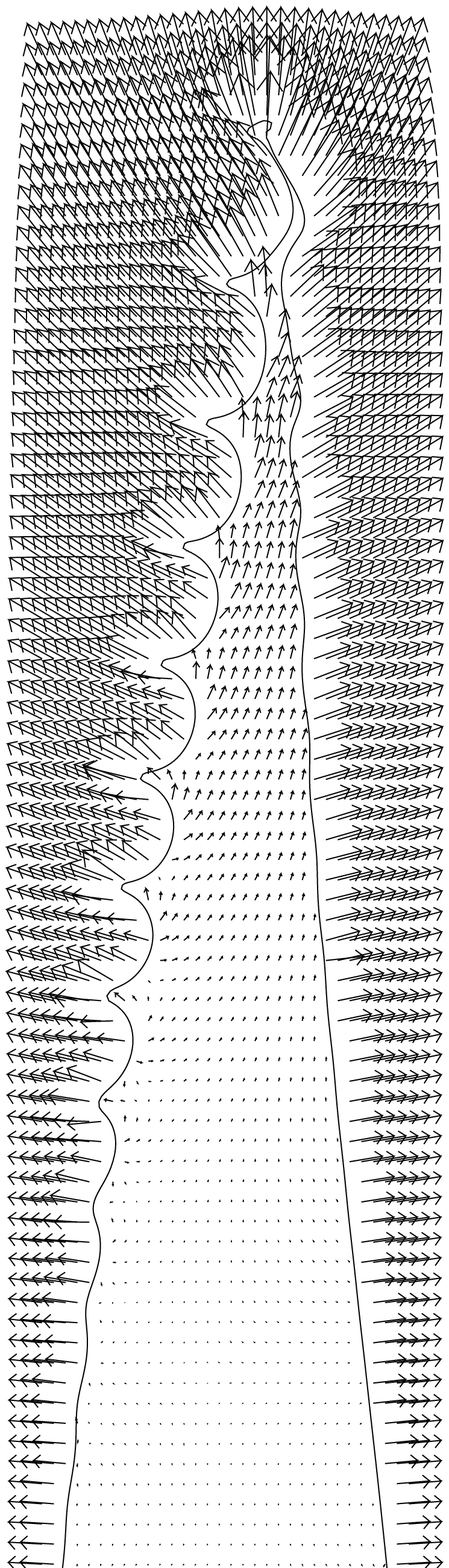}} \par}
\end{figure}

Figure \ref{fig:fleches}: A flame obtained numerically for a sinusoidal
forcing on one side with the associated induced flow field (including
the flow field of the image front). Parameters \( U=10 \) \( L=4 \)
\( \varepsilon =0.2 \) \( a=1 \) \( \gamma =0.8 \) \( \omega =50 \)

\begin{figure}[p]

\caption{\label{fig:ondule} Denet, Phys. Fluids\protect \\
}

{\centering \includegraphics{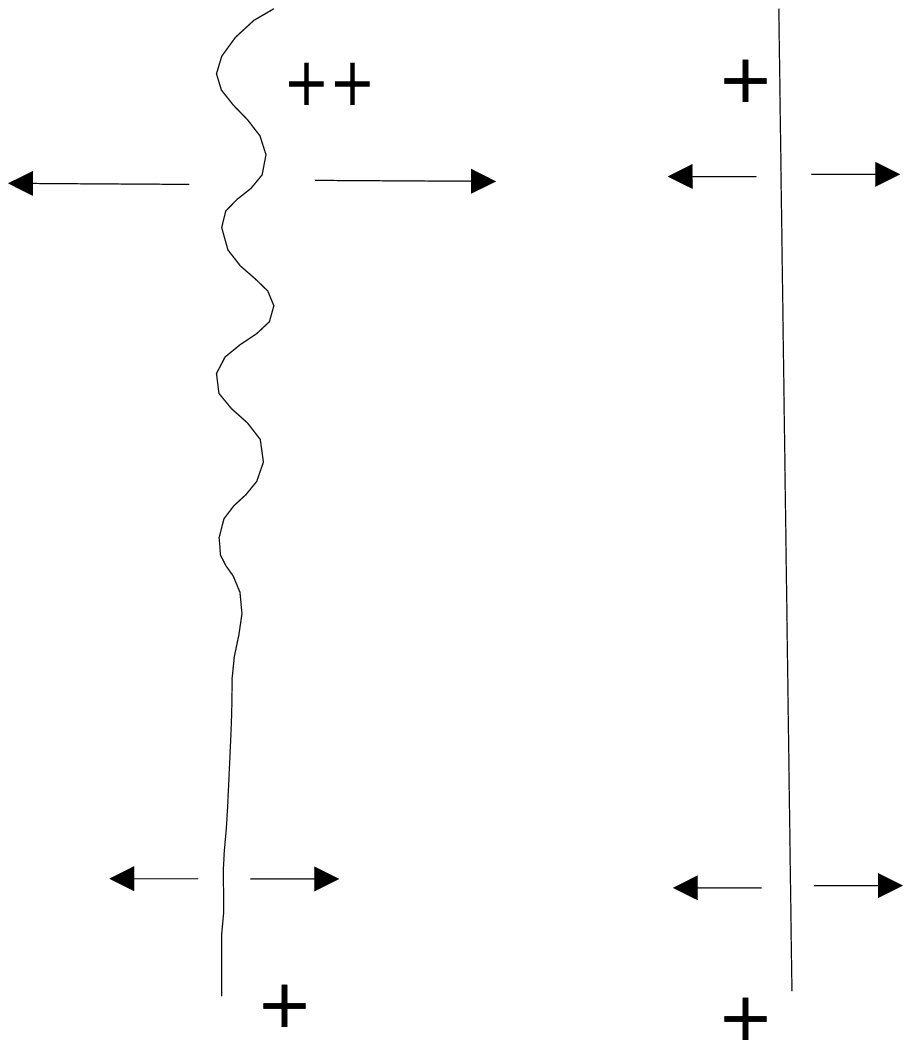} \par}

\vspace{10cm}
\end{figure}

Figure \ref{fig:ondule}: A qualitative explanation of the deviation
observed in the previous figure. Because of Gauss theorem, the perturbed
flame can be seen sufficiently far as a plane with a higher charge.

\begin{figure}[p]

\caption{\label{fig:uinj40aa1g02abblanc5exp08} Denet, Phys. Fluids \protect \\
}

{\centering \includegraphics{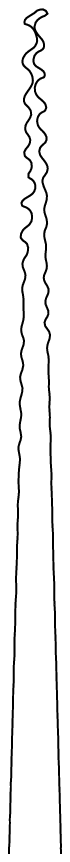} \par}

\vspace{20cm}
\end{figure}

Figure \ref{fig:uinj40aa1g02abblanc5exp08}: Flame excited on both
sides close to the base by a white noise. Parameters \( L=1 \) \( U=40 \)
\( \varepsilon =0.2 \) and \( \gamma =0.8 \), the amplitude of the
white noise being \( a=5 \)

\begin{figure}[p]

\caption{\label{fig:uinj10aa20g02abblanc5exp08} Denet, Phys. Fluids\protect \\
}

{\centering \resizebox*{0.1\textwidth}{!}{\includegraphics{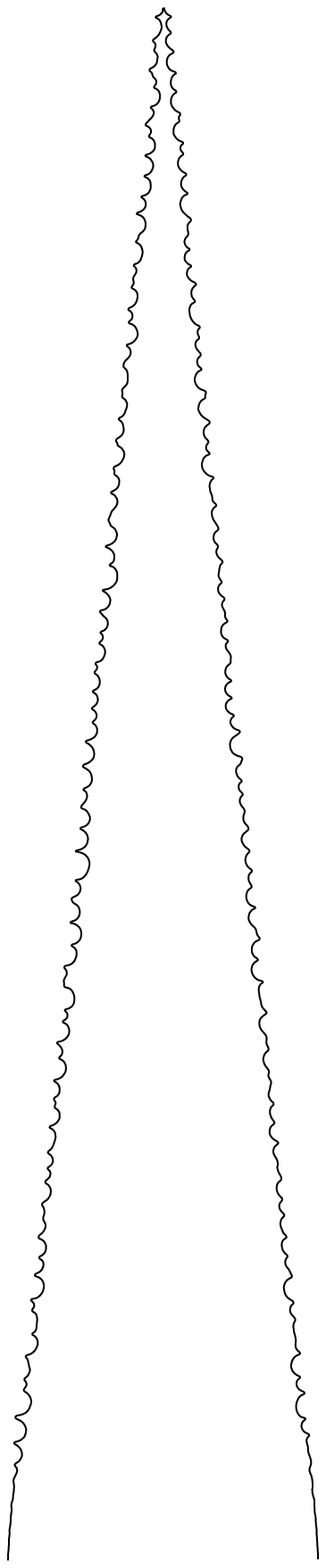}} \par}

\vspace{20cm}
\end{figure}

Figure \ref{fig:uinj10aa20g02abblanc5exp08}: Initial condition for
a simulation with a large domain. The develoment of the perturbations
can be seen in the next figures. Parameters \( U=10 \) \( L=20 \)
\( \varepsilon =0.2 \) \( \gamma =0.8 \)

\begin{figure}[p]

\caption{\label{fig:uinj10aa20g02abblanc0suiteexp08} Denet, Phys. Fluids
\protect \\
}

{\centering \resizebox*{0.1\textwidth}{!}{\includegraphics{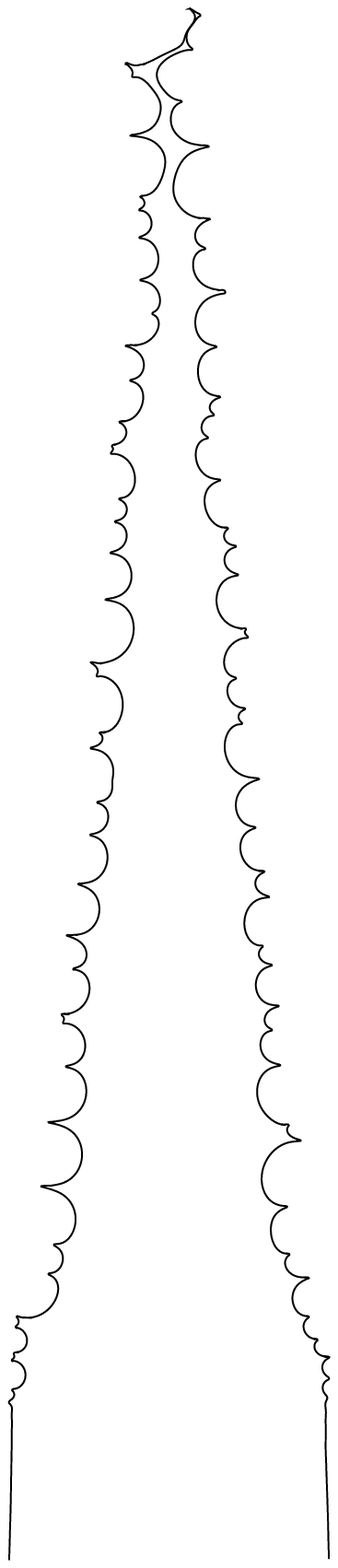}} \par}

\vspace{20cm}
\end{figure}

Figure \ref{fig:uinj10aa20g02abblanc0suiteexp08}: Development of
the instability for the initial condition given in Figure \ref{fig:uinj10aa20g02abblanc5exp08}.

\begin{figure}[p]

\caption{\label{fig:uinj10aa20g02abblanc0suite3exp08} Denet, Phys. Fluids\protect \\
}

{\centering \resizebox*{0.1\textwidth}{!}{\includegraphics{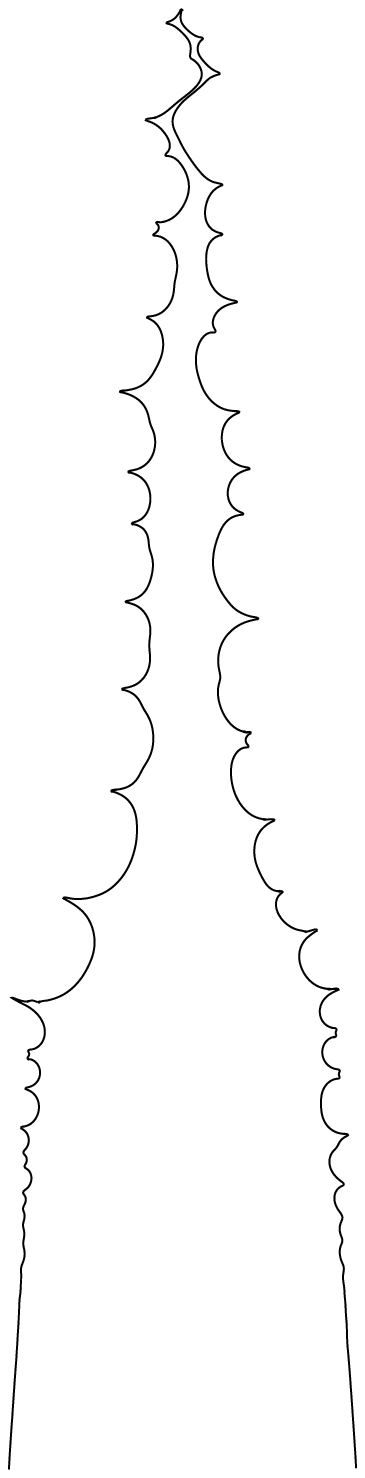}} \par}

\vspace{20cm}
\end{figure}

Figure \ref{fig:uinj10aa20g02abblanc0suite3exp08}: Development of
the instability for the initial condition given in Figure \ref{fig:uinj10aa20g02abblanc5exp08}

\begin{figure}[p]

\caption{\label{fig:uinj10aa20g02abblanc0suite5exp08} Denet, Phys. Fluids
\protect \\
}

{\centering \resizebox*{0.1\textwidth}{!}{\includegraphics{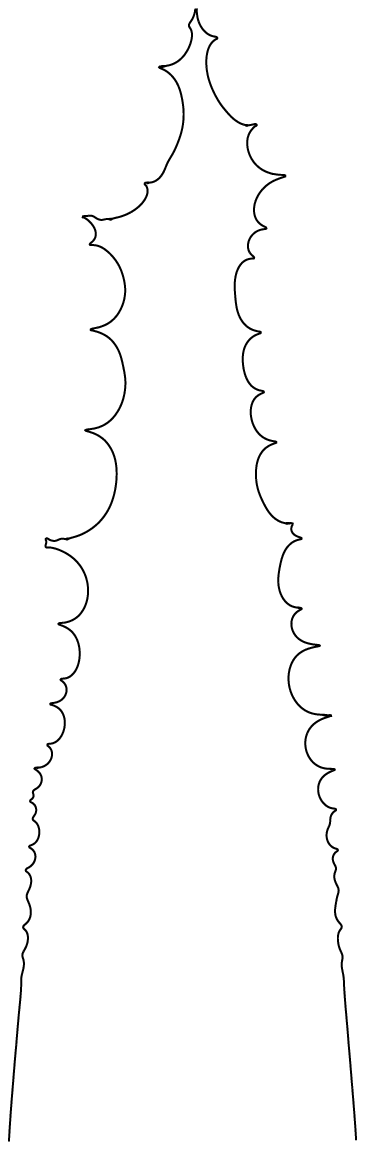}} \par}
\end{figure}

Figure \ref{fig:uinj10aa20g02abblanc0suite5exp08}: Development of
the instability for the initial condition given in Figure \ref{fig:uinj10aa20g02abblanc5exp08}

\end{document}